\documentclass[a4paper,11pt]{article}
\usepackage{jheppub} 
\usepackage{lineno}

\usepackage{physics}
\usepackage{subcaption}
\usepackage{slashed}
\usepackage{layouts}
\usepackage{dsfont}
\usepackage{cancel}

\newcommand{\CME}{C_{\text{CME}}}

\title{On the absence of the Chiral Magnetic Effect in equilibrium QCD}

\author[a]{B. B. Brandt,}
\author[a]{G. Endr\H{o}di,}
\author[a]{E. Garnacho-Velasco,}
\author[a]{and G. Mark\'{o}}

\affiliation[a]{Fakult\"{a}t f\"{u}r Physik, Universit\"{a}t Bielefeld,\\
Universit\"{a}tsstra{\ss}e 25, 33615 Bielefeld, Germany}

\emailAdd{brandt@physik.uni-bielefeld.de}
\emailAdd{endrodi@physik.uni-bielefeld.de}
\emailAdd{egarnacho@physik.uni-bielefeld.de}
\emailAdd{gmarko@physik.uni-bielefeld.de}

\abstract{
In this paper we investigate the chiral magnetic effect (CME): the 
generation of an electric current due to a homogeneous background magnetic
field and a homogeneous chiral imbalance in QCD.
We demonstrate that the leading coefficient describing the CME vanishes
in equilibrium, both for free fermions as well as in full QCD. 
Our full QCD results are based on continuum extrapolated lattice 
simulations using dynamical staggered quarks with physical masses
as well as quenched Wilson quarks.
We show that it is crucial that a gauge invariant ultraviolet regularization is 
used to compute the CME and elaborate on why some of the existing
in-equilibrium calculations of this effect gave a nonzero result.
We stress that our findings imply the absence of a time-independent CME current
flowing in equilibrium QCD, but do not concern the CME as an out-of-equilibrium, 
time-dependent effect.
}

\begin{document}
\maketitle
\flushbottom

\section{Introduction} \label{sec:Introduction}

The vacuum of Quantum Chromodynamics (QCD) is not empty space but a medium filled with virtual particles giving rise to the low-energy facets of the strong interactions. In particular, the topological fluctuations of the gluon fields induce some of the most fascinating aspects of the theory. In specific external environments -- such as background electromagnetic fields or in rotating coordinate systems -- a series of intricate transport phenomena arise, for which the quantum anomaly of QCD plays a special role. These are so-called anomalous transport effects~\cite{Kharzeev:2013ffa,Huang:2015oca,Landsteiner:2016led}.

The most prominent example for anomalous transport phenomena is the Chiral Magnetic Effect (CME), the generation of an electromagnetic current in the presence of a magnetic field and chiral imbalance~\cite{Kharzeev:2007jp,Fukushima:2008xe}. Via the CME, topological gluonic fluctuations, converted to chirality fluctuations through the anomaly, are expected to produce observable signatures in different physical settings. The CME current has been detected in condensed matter systems~\cite{Li:2014bha} and is actively sought for in heavy ion collision experiments~\cite{STAR:2013ksd,STAR:2014uiw,STAR:2021mii}. 
Moreover, the CME has been argued to play a relevant role in astrophysics and cosmology~\cite{Kamada:2022nyt}, e.g.\ for neutrino transport in supernovae~\cite{Yamamoto:2015gzz}.

The CME has been treated within a series of theoretical approaches, including QCD models and effective theories~\cite{Fukushima:2010fe} (see the review~\cite{Miransky:2015ava}), 
holography~\cite{Yee:2009vw,Rebhan:2009vc,Gynther:2010ed},  hydrodynamics~\cite{Sadofyev:2010pr,Shi:2017cpu},
kinetic theory~\cite{Hattori:2019ahi} and 
lattice simulations~\cite{Yamamoto:2011gk,Yamamoto:2011ks,Buividovich:2013hza,Muller:2016jod,Buividovich:2024bmu}, see also the lattice studies looking for indirect effects of the CME~\cite{Buividovich:2009wi,Abramczyk:2009gb,Buividovich:2009my,Braguta:2010ej,Bali:2014vja,Astrakhantsev:2019zkr}.
In this work we will focus on the theoretical aspects of this effect using both lattice and continuum regularizations of the quantum field theory. 
Since the very first study~\cite{Vilenkin:1980fu}, the theoretical understanding of the CME has greatly evolved. 
By now the CME is understood as an out-of-equilibrium effect, for which a time-dependent electric current is induced by a time-dependent chiral imbalance. 
The most general argument for the absence of an in-equilibrium CME comes from Bloch's theorem, which forbids global conserved currents to flow in equilibrium in the thermodynamic limit~\cite{Yamamoto:2015fxa}. However, there are several equilibrium studies in the literature that did predict a non-vanishing CME current. In particular, this includes many of the above listed continuum calculations, based on the static Dirac eigenvalues for free fermions, as well as in-equilibrium lattice simulations~\cite{Yamamoto:2011gk,Yamamoto:2011ks}.
Given the high importance of the CME for a multitude of physical settings, these contradictory findings clearly call for clarification.

In this paper we will determine the CME conductivity in equilibrium from the first principles of QCD. In particular, we will present the first results for the conductivity based on fully interacting in-equilibrium lattice QCD simulations employing physical quark masses and a continuum extrapolation. Our approach generalizes the method that we developed in~\cite{Brandt:2023wgf} for the determination of the conductivity for the chiral separation effect (CSE), an analogous anomalous transport phenomenon~\cite{Son:2004tq,Metlitski:2005pr}. Moreover, we will revisit the calculation of the CME current for free fermions using a gauge-invariant regularization. All our results are compatible with a vanishing CME current in equilibrium.
In turn, we show how regularizations incompatible with the gauge symmetry, more generically, the use of a non-conserved electric current, can erroneously give nonzero results.
We stress that our findings do not concern the out-of-equilibrium effect that arises for time-dependent external fields or chiral imbalance~\cite{Fukushima:2008xe,Fukushima:2010vw,Horvath:2019dvl}.

\section{The Chiral Magnetic Effect in equilibrium} \label{sec:The Chiral Magnetic Effect}

We start our discussion of the CME in thermal equilibrium by describing the setup. We consider a system in the presence of a background magnetic field $B$, where the chiral density is nonzero. The chiral imbalance is parameterized by a chiral chemical potential $\mu_5$, whose detailed properties will be discussed below. The generation of a vector current $J_\nu$ (parallel to $B$) in this scenario is what is known as the chiral magnetic effect. We will be interested in the linear term $\CME$ in a Taylor-expansion of this current for weak $B$ and $\mu_5$, which we will refer interchangeably as CME conductivity coefficient or CME conductivity. Apart from section \ref{sec:cmefree} and unless explicitly stated otherwise, throughout the paper we work in Euclidean space-time and with Hermitian Dirac matrices which satisfy $\{\gamma_\mu,\gamma_{\mu}\}=2\delta_{\mu\nu}$. 

\subsection{Currents and chemical potentials}
\label{sec:curchempot}

We consider the vector current coupled to the electric charge,\footnote{When considering a theory with a single quark flavor, defining the vector current and the chiral chemical potential is unambiguous up to an overall normalization. For several quark flavors, like in full QCD, there are different options for coupling the current and the chiral chemical potential to e.g.\ quark charges or to baryon number. Different choices can in general yield different results for the conductivities, as we showed for example in our previous study of the CSE~\cite{Brandt:2023wgf}. For the CME, this choice is unimportant as we will briefly discuss below in Sec.~\ref{subsec: full qcd}.}
\begin{equation}
J_{\nu}=\sum_f \dfrac{q_f}{e} \,\frac{T}{V}\int \dd^4x \, \bar\psi_f(x) \gamma_\nu \psi_f(x)\,,
\label{eq:veccurdef}
\end{equation}
where $f=u,d,s,\ldots$ labels the quark flavors and $q_f$ are the corresponding electric charges.
The charge density $J_4$ can be controlled by a charge chemical potential in the grand canonical ensemble.
Analogously, the chiral density can also be parameterized by a chiral chemical potential $\mu_5$, which couples to the fourth component of the associated axial current, 
\begin{equation}
    J_{\nu5}=\sum_f \,\frac{T}{V}\int \dd^4x \, \bar\psi_f(x) \gamma_\nu \gamma_5  \psi_f(x)\,,
    \label{eq:axcurdef}
\end{equation}
which we normalized here so that each quark flavor contributes with unit weight.

The chiral chemical potential $\mu_5$ is clearly not a chemical potential in the standard thermodynamic sense, since the charge it couples to is not conserved (due to the nonzero quark mass as well as the chiral anomaly). It can either be thought of as a mere parameter that controls the chiral imbalance of the system or as the constant temporal component of an axial gauge field. In any case, one could be concerned if $\mu_5$ does really induce chiral imbalance or if its effect is washed out due to the non-conserved nature of the axial charge. Below we will demonstrate, both for free quarks and for full QCD, that $\mu_5$ indeed induces nonzero chirality in the system.

The CME conductivity coefficient is defined via the leading-order term in the Taylor-expansion of the electric current with respect to $B$ and $\mu_5$ around vanishing fields,
\begin{equation}
    \label{eq:cme}
    \langle J_{3} \rangle=\CME\, \mu_5 \, e B + \mathcal{O}(\mu_5^3, B^3)\,,
\end{equation}
where we assumed, without loss of generality, that the background magnetic field points in the third spatial direction.
Moreover, the magnetic field is considered in units of the elementary electric charge $e>0$, so that we may work with the renormalization group invariant combination $eB$.

Although full lattice QCD simulations do not suffer from a sign problem at finite $\mu_5$, the introduction of this parameter for staggered fermions leads to technical difficulties which will be discussed below. Instead, we consider the Taylor-expansion of the current expectation value in the chiral chemical potential, evaluated at non-zero external magnetic fields but vanishing $\mu_5$. This technique will also be used for the crosschecks with Wilson fermions. We note that simulations with Wilson fermions at $\mu_5\neq0$ are free of the difficulties present in the staggered formulation and have already been used to investigate the CME~\cite{Yamamoto:2011ks,Yamamoto:2011gk}. Following Eq.~\eqref{eq:cme}, the first $\mu_5$ derivative of the CME current is given by
\begin{equation}
   \eval{\pdv{\langle J_{3} \rangle}{\mu_5}}_{\mu_5=0}= C_{\text{CME}} \, e B\,.
   \label{eq:j35der}
\end{equation}
Using the results for this derivative for a number of different magnetic fields, $\CME$ can be extracted via numerical differentiation with respect to $eB$. Alternatively, using the technique developed in Ref.~\cite{Bali:2015msa}, the derivative with respect to the magnetic field can be carried out analytically. This results in an expression including a three-point function, see Refs.~\cite{Bali:2020bcn,Buividovich:2021fsa} which, at leading order, is a special realization of the triangle diagram in the Adler-Bell-Jackiw anomaly. This shows the tight relation between the conductivity and the anomaly. 

There is an overall proportionality constant involving the quark baryon numbers, the quark charges and the number of colors $N_c=3$, which can be factored out of the conductivity. It reads
\begin{align}
C_{\rm dof} = N_c\sum_f\left(\frac{q_f}{e}\right)^2 \,.
\end{align}
This overall factor can always be restored, and from this point we rescale all our results by $C_{\text{dof}}$ unless explicitly stated otherwise.

\subsection{Analytic result for the CME conductivity for free fermions}
\label{sec:cmefree}

The in-equilibrium effect may be studied in the free case by constructing the electric current operator using the exact Dirac eigenvalues at $B\neq0$ and $\mu_5\neq0$~\cite{Fukushima:2008xe,Sheng:2017lfu}. In these works, the conductivity coefficient was found to be given by $\CME=1/(2\pi^2)$.
However, there is a critical issue related to the ultraviolet regularization, which we will illustrate below.

Let us consider one colorless fermion flavor (with mass $m$ and charge $q$) at zero temperature $T$. Instead of working directly with the Dirac eigenvalues, we proceed as in Eq.~\eqref{eq:j35der}, i.e.\ we will evaluate the $\mu_5$-derivative of the vector current in a homogeneous magnetic field background, for which free fermion propagators are known exactly~\cite{Schwinger:1951nm, Shovkovy:2012zn}. Unlike in the rest of the text, we will carry out the calculation using the Minkowski metric. Accordingly, the Dirac matrices are the Minkowskian ones fulfilling $\{\gamma_\mu,\gamma_\nu\}=2\eta_{\mu\nu}=2\,\textmd{diag}(1,-1,-1,-1)$.

The chiral chemical potential $\mu_5$ couples to the four-volume integral of the zeroth component of the axial current, therefore the $\mu_5$-derivative of the third component of the vector current can be written as
\begin{align}
    \left.\frac{\partial\langle J_{3}\rangle}{\partial\mu_5}\right|_{\mu_5=0}
    =  \frac{T}{V}
    \int \dd^4x \int \dd^4y \,\langle \bar\psi(x)\gamma_3\psi(x)\bar\psi(y)\gamma_0\gamma_5\psi(y)\rangle\,.
    \label{eq:A1}
\end{align}
We regularize in the ultraviolet using the Pauli-Villars (PV) scheme, meaning that all diagrams are replicated by the regulator fields with coefficients $c_s$ and masses $m_s$. The details of the regularization are text-book knowledge found e.g.\ in Ref.~\cite{Itzykson:1980rh}. We recall here that we need three extra fields altogether, and reserving $s=0$ for the physical field (with physical mass $m$) the parameters will be
\begin{align}
    c_0&=c_1=1\,,\quad c_2=c_3=-1\,,\\
    m_0^2&=m^2\,,\quad m_1^2=m^2+2\Lambda^2\,,\quad m_2^2=m_3^2=m^2+\Lambda^2\,,
    \label{eq:PVmasses}
\end{align}
with $\Lambda\to\infty$ to be taken at the end of the calculation. 

The derivative according to Eq.~\eqref{eq:j35der} is proportional to the CME coefficient, and using Wick's theorem we can rewrite the right-hand side of Eq.~\eqref{eq:A1} to obtain
\begin{align}
    \label{eq:cse_PV_start}
    \CME \, qB = \frac{iT}{V}\sum_{s=0}^{3} c_s\int \dd^4 x\int \dd^4y \Tr\left[\gamma_3 S_s(x,y) \gamma_0 \gamma_5 S_s(y,x)\right]\,,
\end{align}
where the PV fields are already taken into account, and $S_s$ is the fermion propagator for the field $s$ in a homogeneous magnetic background. The latter reads~\cite{Shovkovy:2012zn}
\begin{align}
S_s(x,y) = \Phi(x,y)\int\frac{\dd^4p}{(2\pi)^4}{\rm \,e\,}^{-ip(x-y)}\widetilde S_s(p)\,.
\end{align}
Here 
\begin{align}
\Phi(x,y) = \exp\left[iqB(x_1+y_1)(x_2-y_2)/2\right]\,,
\end{align}
is the Schwinger phase, and
\begin{align}
\widetilde S_s(p) = \int_0^\infty \dd z {\rm \,e\,}^{iz(p_0^2 - m_s^2-p_3^2) - i\frac{p_1^2+p_2^2}{|qB|}\tan{(z|qB|)}}&\left[\slashed{p}+m_s+(p_1\gamma_2-p_2\gamma_1)\tan(zqB)\right]\nonumber\\
&\times\left[1-\gamma_1\gamma_2\tan(zqB)\right]\,.
\end{align}
Carrying out the traces 
\begin{align}
\label{eq:cmetan}
    \CME \,qB=-4\sum_{s=0}^{3} c_s &\int\frac{\dd^4p}{(2\pi)^4}\int_0^\infty \dd z_1\, \dd z_2 {\rm \,e\,}^{i(z_1+z_2)(p_0^2 - m_s^2-p_3^2)}\left(m_s^2-p_0^2-p_3^2\right) \nonumber\\
    &{\rm \,e\,}^{- i\frac{p_1^2+p_2^2}{qB}[\tan{(z_1qB)}+\tan{(z_2qB)}]}\left[\tan(z_1qB)+\tan(z_2qB)\right]\,.
\end{align}
Now we Wick rotate the momentum components and the Schwinger parameters, $p_0 = ip_4\,,iz_1 = Z_1\,,iz_2 = Z_2$. The $p_1$ and $p_2$ integrals are simple Gaussians, and evaluating them factorizes the magnetic field dependence, uncovering an explicitly linear behavior in $B$. This allows us to cancel the $qB$ factor on both sides, resulting in
\begin{align}
     \CME=\frac{1}{4\pi^3}\sum_{s=0}^{3} c_s\int_{-\infty}^\infty \dd p_3 \,\dd p_4 \int^\infty_0 \dd Z_1 \dd Z_2 {\rm \,e\,}^{-(Z_1+Z_2)(p_4^2 + m_s^2+p_3^2)}\left(m_s^2+p_4^2-p_3^2\right)\,.
     \label{eq:cme_p3z1z2}
\end{align}
Carrying out the integrals over $Z_1$ and $Z_2$, we arrive at the final formula
\begin{align}
\label{eq:cmecancel}
    \CME=\frac{1}{4\pi^3}\sum_{s=0}^{3} c_s\int_{-\infty}^\infty \dd p_3 \,\dd p_4 \frac{m_s^2+p_4^2-p_3^2}{(m_s^2+p_4^2+p_3^2)^2} 
    \propto \sum_{s=0}^3 c_s = 0\,.
\end{align}
The integral for a single PV field is a peculiar improper integral. On dimensional grounds, its result does not depend on $m_s$, and hence the sum over the PV fields inevitably vanishes due to~\eqref{eq:PVmasses}, as we indicated in~\eqref{eq:cmecancel}. However, to obtain a specific value for just one of the PV fields, one must choose a certain parameterization of the $p_3-p_4$ plane spanned by the two integration variables. As we show and analyze in App.~\ref{app:RegDep}, the result of the integration depends on the parameterization, reflecting that the unregularized result is physically meaningless. The regularized result, in turn, is $\CME=0$, irrespectively of the parameterization.

This issue with regularization has already been discussed in the literature \cite{Buividovich:2013hza,Buividovich:2024bmu,Hou:2011ze,Zubkov:2016tcp,Horvath:2019dvl}. In App.~\ref{sec:cmefinT} we show a generalization of this calculation to finite temperature, where $\CME$ is found to vanish as well.

Another way of understanding this cancellation in equilibrium is through Bloch's theorem~\cite{PhysRev.75.502}. This no-go theorem from quantum mechanics forbids conserved currents to flow in equilibrium in the thermodynamic limit. The theorem can be generalized to quantum field theory and applied to the CME~\cite{Yamamoto:2015fxa}. Since the electromagnetic current is a conserved current, Bloch's theorem forbids a non-zero expectation value in equilibrium and the CME has to be absent. We also note that this argument does not apply to the chiral separation effect (CSE)~\cite{Son:2004tq,Metlitski:2005pr}, where a non-conserved axial current is generated in a dense magnetized medium.

We emphasize here that Bloch's theorem indicates the absence of the current expectation value at any non-zero $\mu_5$ and $B$ (in sufficiently large volumes). Consequently, while focusing on the leading coefficient $\CME$, we expect all coefficients in the Taylor expansion of Eq.~\eqref{eq:cme} to vanish identically.

\subsection{Chiral imbalance}
A final remark in this section concerns the relation between the chiral chemical potential and the chiral imbalance. We emphasize that the absence of CME in equilibrium is not due to the absence of chirality in the system, but it arises from a non-trivial cancellation as shown in Eq.~\eqref{eq:cmecancel}. In fact, it can be shown that $\mu_5$ does induce nonzero chiral density. This can be checked by calculating the axial density $J_{45}$ of Eq.~\eqref{eq:axcurdef} in the presence of a (baryon) chiral chemical potential.

However, later we will check the presence of chirality using lattice simulations, and, as already mentioned, we avoid simulations at non-zero $\mu_5$. Instead, we consider the coefficient of the leading order term in the expansion of $\langle J_{45}\rangle$ in $\mu_5$. The coefficient is equivalent to the second derivative of the partition function with respect to $\mu_5$, i.e.\ the axial susceptibility
\begin{equation}
    \chi^b_5(T)=\dfrac{T}{V}\eval{\pdv[2]{\log \mathcal{Z}}{\mu_5}}_{\mu_5=0}\, .
    \label{eq:axialsusc}
\end{equation}
Its nonzero value at $\mu_5=0$ indicates that a weak chiral chemical potential induces nonzero chiral imbalance.
As we show in App.~\ref{sec:freechi5}, the susceptibility is a bare observable, subject to additive renormalization, which we indicate by the superscript $b$. The divergent part can be removed by subtracting the contribution at $T=0$,
\begin{equation}
    \chi_5(T)=\chi^b_5(T)-\chi^b_5(T=0)\, .
    \label{eq:chi5sub}
\end{equation}
For non-interacting fermions, the renormalized axial susceptibility takes the form
\begin{equation}
\label{eq:chi5analy}
      \dfrac{\chi_5}{T^2}=\frac{4}{\pi^2}\int^\infty_0 \dd p\, \dfrac{p^2}{\qty(1+e^{\sqrt{(m/T)^2+p^2}})\,\sqrt{(m/T)^2+p^2}} \quad\xrightarrow{m/T\rightarrow0}\quad \dfrac{1}{3}\,.
\end{equation}
The calculation can again be found in App.~\ref{sec:freechi5}.

\begin{figure}[ht]
    \centering
    \includegraphics{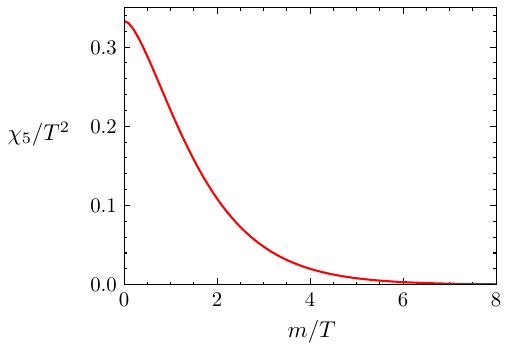}
    \caption{Analytic result (\ref{eq:chi5analy}) for the axial susceptibility $\chi_5$ for non-interacting fermions as a function of $m/T$. }
    \label{fig:chi5_analytic}
\end{figure}

\section{Lattice methods} \label{Simulation setup}

In the following, we will use lattice simulations to check the above expectations in full QCD. In addition, our aim is to shed light on former lattice results where a non-zero $\CME$ has been observed~\cite{Yamamoto:2011ks,Yamamoto:2011gk}. As will become clear, the reason for this apparent discrepancy with the expectations is the use of a non-conserved vector current in these studies.
The appearance of a conserved current in the CME is the reason for the applicability of Bloch's theorem to conclude that the CME should be absent in equilibrium. Retaining the conserved current in lattice simulations turns out to be vital to obtain consistent results.

We concentrate on the Euclidean lattice formulation of the chiral magnetic effect. In order to define the lattice observables required for the calculation of $\CME$, we first need to describe how the magnetic field and the chiral chemical potential enter the discretized theory. The magnetic field $B$ is taken to be constant and homogeneous, pointing in the third spatial direction. As already mentioned above, we will work with the renormalization group invariant combination $eB$. 

Considering the relation between the Minkowski (indicated by the superscript $\rm M$) and Euclidean Dirac matrices ($\gamma_0^{\rm M}=\gamma_4$, $\gamma_i^{\rm M}=i\gamma_i$), we find that the Minkowski-space observable from Eq.~\eqref{eq:j35der} is given by
\begin{equation}
   \Re \eval{\pdv{\langle J_{3}\rangle}{\mu_5}}_{\mu_5=0}^{\rm M} = -\Im \eval{\pdv{\langle J_{3}\rangle}{\mu_5}}_{\mu_5=0}
   \label{eq:EuclMink}
\end{equation}
in terms of Euclidean operators. The latter will be considered in the rest of the manuscript. We also checked explicitly that the real part of the Euclidean current expectation value is consistent with zero.

\subsection{Lattice setup -- staggered quarks}

We discuss first the setup with rooted staggered fermions, which we use to study the CME in full QCD with $2+1$ flavors at the physical point. In this formalism, the partition function $\mathcal{Z}$ can be written using the Euclidean path integral over the gluon links $U$ 
\begin{equation}
    \mathcal{Z}=\int \mathcal{D} U \exp[-\beta S_g] \,\prod_f \qty[ \det M_f(U,q_f,m_f)]^{1/4},
    \label{eq:partfunc}
\end{equation}
where the fermionic fields have already been integrated out to yield the fermion determinant, $\beta=6/g^2$ denotes the inverse gauge coupling and $m_f$ are the quark masses for each flavor $f=u,d,s$. In Eq.~\eqref{eq:partfunc}, $S_g$ is the gluonic action and $M_f$ is the massive staggered Dirac operator, which contains the quark charges $q_u/2=-q_d=-q_s=e/3$. For $S_g$, we used a tree-level Symanzik action, and the Dirac operator contains twice stout-smeared links, which reduce the effect of the unphysical splitting of the masses of the different taste degrees of freedom in the staggered formulation.

The quark masses are tuned to the physical point as a function of the lattice spacing $a$ along the line of constant physics (LCP)~\cite{Borsanyi:2010cj}. The chiral chemical potential (just like the quark mass) has a non-trivial multiplicative renormalization constant -- and also induces divergent terms in $\log\mathcal{Z}$ as we discuss in App.~\ref{sec:freechi5}. However, unlike the mass renormalization constant $Z_S^{-1}$, the relevant (flavor-singlet) axial-vector renormalization constant $Z_A^{-1}$ arises from a two-loop diagram and approaches unity perturbatively in the continuum limit~\cite{Constantinou:2016ieh,Bali:2021qem}. Thus, we expect that its use would only affect discretization errors in our lattice simulations and we do not include such a multiplicative renormalization for our observables, neither an effect of $\mu_5$ in our LCP. We mention that such effects would in any case enter at $\mathcal{O}(\mu_5^2)$ due to charge conjugation symmetry, and therefore do not impact our observable~\eqref{eq:j35der}, which contains a first derivative with respect to $\mu_5$ at $\mu_5=0$. The simulations are performed on a four-dimensional lattice with $N_s$ spatial and $N_t$ temporal points, each site $n$ labeled by four integers $n=\{n_1,n_2,n_3,n_4\}$. The physical spatial volume is given by $V=L^3=(aN_s)^3$ and the temperature by $T=(aN_t)^{-1}$. Also the unitary vector in the direction $\nu$ is denoted by $\hat{\nu}$.

We note that the magnetic field is included in our setup as a classical background field, so we do not consider dynamical photons in our calculations. Analogously to the $\mathrm{SU}(3)$ links, the electromagnetic potential enters the Dirac operator for the quark flavor $f$ as a parallel transporter between two lattice points $u_{f\nu}=\exp(iaq_fA_\nu)$. These $\mathrm{U}(1)$ links are chosen such that they represent a homogeneous magnetic field in the $x_3$ direction. The periodic boundary conditions in our finite volume imply that the flux of the field is quantized as $eB=6\pi N_b/(aN_s)^2$ with the integer flux quantum $N_b\in\mathds{Z}$~\cite{Bali:2011qj}. 

The staggered transformation partially diagonalizes the Dirac operator, by mixing coordinate and Dirac indices, so that the quark fields $\psi_f$ are transformed to the so-called staggered fields $\chi_f$. 
It is possible to find a conserved vector current and an anomalous (flavor-singlet) axial current in the staggered formalism, via the point-split bilinears~\cite{Sharatchandra:1981si},
\begin{equation}
    J_\nu^f(n) = \Bar{\chi}_f(n)  \Gamma^f_\nu(n,m) \chi_f(m)\,, \qquad
        J^f_{\nu5}(n) = \Bar{\chi}_f(n)  \Gamma^f_{\nu5}(n,m) \chi_f(m)\,,
\end{equation}
involving the staggered counterparts of the Dirac matrices~\cite{Durr:2013gp},
  \begin{align}
  \label{eq:gammas}
    \Gamma^f_\nu(n,m)&=\dfrac{\eta_\nu(n)}{2}\left[U_\nu(n)u_{f\nu}(n) \,e^{h(\mu_5)}\delta_{n+\hat{\nu},m}+U^{\dagger}_\nu(n-\hat{\nu})u^{*}_{f\nu}(n-\hat{\nu}) \,e^{-h(\mu_5)}\delta_{n-\hat{\nu},m}\right], \nonumber \\
        \Gamma^f_{\nu5} &=\dfrac{1}{3!}\sum_{\rho,\alpha,\beta} \epsilon_{\nu\rho\alpha\beta}\,\Gamma^f_\rho\Gamma^f_\alpha\Gamma^f_\beta\,.
    \end{align}
The $\Gamma_\nu^f$ objects also enter the staggered equivalent of the fifth Dirac matrix,
\begin{equation}
                \Gamma^f_{5} =\dfrac{1}{4!}\sum_{\nu,\rho,\alpha,\beta} \epsilon_{\nu\rho\alpha\beta}\,\Gamma^f_\nu\Gamma^f_\rho\Gamma^f_\alpha\Gamma^f_\beta\,.
                \label{eq:gammas2}
\end{equation}

Above, $\eta_\nu(n)=(-1)^{\sum_{\rho<\nu}n_\rho}$ are the staggered phases
and $\epsilon_{\nu\rho\alpha\beta}$ the totally antisymmetric four-index tensor with the convention
$\epsilon_{1234}=+1$. These Dirac structures depend explicitly on the chiral chemical potential via the factor $e^{\pm h(\mu_5)}$. The exponential way of introducing the chiral chemical potential is analogous to the case of a baryonic one~\cite{Hasenfratz:1983ba}, which ensures the correct continuum properties of the corresponding currents. To linear order in $\mu_5$, the functional form can be found by noting that 
\begin{equation}
    \eval{\pdv{M^f}{\mu_{5}}}_{\mu_5=0}= \Gamma^f_{45}\,,
    \label{eq:mu5introductionMf}
\end{equation}
leading to 
\begin{equation}
\label{eq:hmu5}
    h(\mu_5)=a\mu_{5}\,\Sigma^f_\nu(\mu_5)\,,
\end{equation}
with the staggered representations of the spin operator,
\begin{align}
    &\Sigma^f_\nu(\mu_5)=\frac{1}{3!}\epsilon_{\nu\rho\alpha4}\Gamma^f_\rho\Gamma^f_\alpha(\mu_5)\,.
\end{align}

Altogether, the Dirac operator in the presence of a chiral chemical potential in the staggered formalism can be written as (notice that $\Sigma_4^f=0$),
\begin{equation}
\begin{split}
    \slashed{D}^f(n,m)=\dfrac{1}{2a}\sum_{\nu=1}^4\eta_\nu(n)\Big[&U_\nu(n)u_{f\nu}(n)\,e^{a\mu_{5} \Sigma^f_\nu(\mu_5)}\delta_{m,n+\hat\nu}\\
    -&U^\dagger_\nu(n-\hat{\nu})u^*_{f\nu}(n-\hat{\nu})\,e^{-a\mu_{5} \Sigma^f_\nu(\mu_5)}\delta_{m,n-\hat\nu}\Big]\,.
\end{split}    
    \label{eq:diracopstag}
\end{equation}
Here we note the difficulties of simulating at non-zero $\mu_5$ with staggered fermions. The Dirac operator (\ref{eq:diracopstag}) has a recursive dependence on $\mu_5$, since each spatial hopping involves the $\Sigma_i^f$ operator, which itself contains spatial hoppings. Even if this dependence is truncated at a certain order, the Dirac operator becomes a highly non-local operator involving the exponentials of the $\Sigma_i^f$ matrices. An alternative approach is to introduce the chiral chemical potential in a linear way, see \cite{Braguta:2015zta, Astrakhantsev:2019wnp,Braguta:2015owi}. Both formulations lead to the same expressions for first derivatives like the CME current~\eqref{eq:j35der}, however they differ for higher derivatives like the axial susceptibility~\eqref{eq:axialsusc}. We get back to the latter point below in Sec.~\ref{subsec: full qcd}. 

With these definitions, we can now perform the Grassmann integral over the staggered fields, giving the expectation value for the (volume-averaged) vector current,
\begin{equation}
 \langle J_{3} \rangle = 
 \frac{T}{V} \frac{1}{4}\sum_f \frac{q_f}{e}\left\langle \Tr \left(\Gamma^f_{3}M_f^{-1}\right)  \right\rangle\,,
\end{equation}
where $\Tr$ refers to a trace in color space and a summation over the lattice coordinates. The factor $1/4$ results from rooting. Now we can calculate the derivative~\eqref{eq:j35der}, required to extract $\CME$. The obtained observable is very similar to the one used to calculate the CSE conductivity~\cite{Brandt:2023wgf}. It contains the usual disconnected and connected terms, together with an additional tadpole term arising due to the derivative of $\Gamma^f_{3}$ with respect to $\mu_5$, 
\begin{equation}
\begin{split}
   C_{\rm CME} \,eB = \eval{\pdv{\langle J_{3}\rangle}{\mu_5}}_{\mu_5=0}=\dfrac{T}{V}\Bigg[\dfrac{1}{16}&\sum_{f,f'}\frac{q_f}{e}\expval{\text{Tr}\qty(\Gamma^{f'}_{45} M_{f'}^{-1})\text{Tr}\qty(\Gamma^{f}_{3} M_{f}^{-1})}\\
    -\dfrac{1}{4} &\sum_f \frac{q_f}{e}\expval{\text{Tr}\qty(\Gamma^f_{45} M_f^{-1}\Gamma^f_{3} M_f^{-1})}\\
    +\dfrac{1}{4}&\sum_f  \frac{q_f}{e}\expval{\text{Tr}\qty(\dfrac{\partial\Gamma^f_{3}}{\partial \mu_{5}}M_f^{-1})}\Bigg]\,.
    \end{split} \label{eq:derstag}
\end{equation}
 These expectation values are to be evaluated at $\mu_5=0$ but nonzero magnetic field $eB$. 

One of our most important goals is to compare with the non-vanishing CME results with Wilson fermions~\cite{Yamamoto:2011gk,Yamamoto:2011ks}, both in full QCD and in the quenched approximation. To enable a direct comparison, we also consider QCD in the quenched approximation. 
This approximation corresponds to dropping the fermion determinant from the partition function~\eqref{eq:partfunc}, while leaving the fermionic operators in the observables unchanged. This results in a non-unitary theory where quarks behave differently in operators as in loop diagrams. It is a commonly employed approximation that simplifies simulation algorithms significantly, although is only exact for infinite quark masses.
For completeness, in the quenched case we employ both the staggered and the Wilson discretization of fermions. 

Finally, we also calculate $\CME$ in the absence of gluonic interactions, both for Wilson and for staggered quarks. To this end, we calculated the eigenvalues and eigenvectors of the staggered Dirac operator exactly and constructed the necessary traces from these (see App.~\ref{sec: free appendix}), while for Wilson fermions we relied on stochastic techniques to estimate the traces.

Above we also introduced the axial susceptibility $\chi_5$, which measures the linear response of the chiral density to the introduction of $\mu_5$. It is an important cross-check that $\chi_5$ is non-zero in our system, since the absence of chirality would automatically imply a vanishing CME. In the staggered formulation, this observable reads 
\begin{equation}
\begin{split}
   \dfrac{\chi_5}{T^2}=\dfrac{1}{TV}\Bigg[\dfrac{1}{16}&\sum_{f,f'}\expval{\text{Tr}\qty(\Gamma^f_{45} M_f^{-1})\text{Tr}\qty(\Gamma^{f'}_{45} M_{f'}^{-1})}
    -\dfrac{1}{4} \sum_f\expval{\text{Tr}\qty(\Gamma^f_{45} M_f^{-1}\Gamma^f_{45} M_f^{-1})}\\
    +\dfrac{1}{4}&\sum_f \expval{\text{Tr}\qty(\dfrac{\partial^2 M_f }{\partial \mu^2_{5}}M_f^{-1})}\Bigg]\,,
    \end{split} \label{eq:chi5stag}
\end{equation}
where again a tadpole term is present.

\subsection{Lattice setup -- Wilson quarks}
Next we describe our setup involving Wilson quarks. This discretization we only consider for free fermions and in the quenched approximation -- simulations with dynamical non-degenerate light Wilson quarks (differing in their electric charges) would be a computationally much more challenging task.

The massive Dirac operator is  
\begin{equation}
    aM^f = C(1-\kappa H)
\end{equation}
with the hopping operator 
\begin{align*}
    H= \sum_n \sum_\nu \Big[&(r-\gamma_\nu\,e^{a\mu_{5}\delta_{\nu4}})U_\nu(n)u_{f\nu}(n) \delta_{n+\hat{\nu},m} \\
    &+(r+\gamma_\nu\,e^{-a\mu_{5}\delta_{\nu4}})U_\nu^\dagger(n-\hat{\nu})u^*_{f\nu}(n-\hat{\nu}) \delta_{n-\hat{\nu},m}\Big]
\end{align*}
and
\begin{align}
    C = am_f+4\,, \qquad
    \kappa = \dfrac{1}{2am_f+8}.
\end{align}
The factor $r$ is the coefficient of the usual Wilson term (taken to be unity in our setup) and $\kappa$ is the hopping parameter.

We note that the chiral chemical potential does not couple to the Wilson term, since the topological part of the axial anomaly for Wilson fermions arises from terms proportional to $r$, and thus the anomalous axial current, to which $\mu_5$ couples, should not contain these terms~\cite{Karsten:1980wd}. In other words, this is the equivalent of~\eqref{eq:mu5introductionMf} for Wilson fermions.

In this formulation, the currents are constructed from the bispinor fields $\psi_f$,
\begin{equation}
    J_\nu^f(n) = \Bar{\psi}_f(n)  \Gamma^f_\nu(n,m) \psi_f(m)\,, \qquad
        J^f_{\nu5}(n) = \Bar{\psi}_f(n)  \Gamma^f_{\nu5}(n,m) \psi_f(m)\,.
\end{equation}
The conserved vector and anomalous axial currents again involve a point-splitting, just as for staggered quarks~\cite{Karsten:1980wd}
\begin{align*}
         \Gamma^f_\nu(n,m)&=\frac{1}{2}\Big[(\gamma_\nu e^{a\mu_{5}\delta_{\nu4}}-r)U_\nu(n) u_{f\nu}(n) \, \delta_{m,n+\hat{\nu}} \\
         &\qquad\quad+(\gamma_\nu e^{-a\mu_{5}\delta_{\nu4}}+r)U^\dagger_\nu(n-\hat{\nu})u^*_{f\nu}(n-\hat{\nu})\, \delta_{m,n-\hat{\nu}}\Big]\,,\\
            \Gamma^f_{\nu5}(n,m)&=\frac{1}{2}\Big[\gamma_\nu\gamma_5 U_\nu(n) u_{f\nu}(n) \,e^{a\mu_{5}\delta_{\nu4}} \delta_{m,n+\hat{\nu}}\\
	&\qquad\quad+\gamma_\nu\gamma_5 U^\dagger_\nu(n-\hat{\nu}) u^*_{f\nu}(n-\hat{\nu})\,e^{-a\mu_{5}\delta_{\nu4}}\delta_{m,n-\hat{\nu}}\Big]\,.
\end{align*}
 
Using the above definitions, the expectation value of the vector current reads,
\begin{equation} 
 \langle J_{3} \rangle = 
 \frac{T}{V} \sum_f \frac{q_f}{e}\left\langle \Tr \left(\Gamma^f_{3}M_f^{-1}\right)  \right\rangle\,.
\end{equation}
In this formalism, $\Gamma^f_3$ only involves hoppings in the third direction, and therefore it is independent of $\mu_5$. Thus, the tadpole term is absent and our observable takes a simpler form than in the staggered case, 
\begin{equation}
\begin{split}
   \CME \, eB = \eval{\pdv{\langle J_{3}\rangle}{\mu_5}}_{\mu_5=0}=\dfrac{T}{V}\Bigg[&\sum_{f,f'}\frac{q_f}{e}\expval{\text{Tr}\qty(\Gamma^{f'}_{45} M_{f'}^{-1})\text{Tr}\qty(\Gamma^{f}_{3} M_{f}^{-1})}\\
    -&\sum_f\frac{q_f}{e}\expval{\text{Tr}\qty(\Gamma^f_{45} M_f^{-1}\Gamma^f_{3} M_f^{-1})} \Bigg]\,,
\end{split}
\label{eq:derwil}
\end{equation}
involving only disconnected and connected terms. 

It is also possible to consider a local vector current
\begin{equation}
J_{\nu}^{f,\rm loc}(n) = \bar\psi_f(n) \gamma_\nu \psi_f(n)
\label{eq:localveccur}
\end{equation}
instead. Even though it is not conserved, it has the same quantum numbers as $J^f_\nu$ and is often employed in Wilson fermion simulations, for example for the study of various aspects of hadron physics and, what is more relevant for this work, it was also used to study the CME in quenched and dynamical QCD~\cite{Yamamoto:2011gk,Yamamoto:2011ks}. We can introduce a chiral chemical potential $\mu_5^{\rm loc}$ that couples to these local currents in the action.
The analogue of~\eqref{eq:derwil} now reads 
\begin{equation}
\begin{split}
    \CME^{\rm loc} \, eB = \eval{\pdv{\langle J^{\rm loc}_{3}\rangle}{\mu_5}}_{\mu_5=0}=\dfrac{T}{V}\Bigg[&\sum_{f,f'}\frac{q_f}{e}\expval{\text{Tr}\qty(\Gamma^{f'}_{45} M_{f'}^{-1})\text{Tr}\qty(\gamma_3 M_{f}^{-1})}\\
    -&\sum_f \frac{q_f}{e}\expval{\text{Tr}\qty(\Gamma^f_{45} M_f^{-1}\gamma_3 M_f^{-1})} \Bigg]\,.
\end{split}    
\label{eq:derwilloc}
\end{equation}

This introduction of a chemical potential is a naive extension of the continuum formulation to the discretized theory and it is well known that this leads to new ultraviolet divergences~\cite{Hasenfratz:1983ba}.
Therefore we use this setup with the non-conserved, local current merely for comparison, and we will test the impact of these ultraviolet divergences on the observable~\eqref{eq:derwilloc} below.

The axial susceptibility in this case takes the form 
\begin{equation}
\begin{split}
  \chi_5=\dfrac{T}{V}\Bigg[&\sum_{f,f'}\expval{\text{Tr}\qty(\Gamma^f_{45} M_f^{-1})\text{Tr}\qty(\Gamma^{f'}_{45} M_{f'}^{-1})}
    -\sum_f\expval{\text{Tr}\qty(\Gamma^f_{45} M_f^{-1}\Gamma^f_{45} M_f^{-1})}\\
        +&\sum_f \expval{\text{Tr}\qty(\dfrac{\partial^2 M_f }{\partial \mu^2_{5}}M_f^{-1})}\Bigg]\,,
\end{split}
\label{eq:chi5wil}
\end{equation}
where now a tadpole term appears also in the Wilson formulation, since $\Gamma^f_{45}$ depends on $\mu_5$.

\subsection{Simulation setup} \label{subsec:Simulation setup}
For our numerical calculations, we follow the same approach as in \cite{Brandt:2023wgf}. The traces are estimated with 100 Gaussian noise vectors, except for free staggered fermions where we used the exact eigensystem instead.

$\CME$ is extracted by calculating the current derivative at different values of $eB$, and then performing a linear fit with no constant term, via a usual $\chi^2$ minimization method. The slope of the fitted linear function then corresponds to the CME conductivity. For the estimation of statistical uncertainties, both of the simulation points and the fits, we use the jackknife method with 10 bins. A systematic error for the coefficient is also estimated by repeating the fit successively, eliminating the data point at the largest magnetic field in each iteration, until only one point is left. The maximum difference in the slope between the original fit and the repeated fits we consider to be the systematical error of $\CME$.

For the quenched data, we use a slightly different approach. Since the determinant is taken to be constant, the measurements at different magnetic fields are all performed on the same gauge configurations, and are therefore correlated. To take this into account, we perform correlated fits considering the covariance matrix of the data in the $\chi^2$ minimization method. While the statistical error of these fits is again calculated with the jackknife method, we consider a less conservative estimation of the statistical error in this case. We perform a wide variety of fits by changing the number of points considered, and including fits with a cubic term to the fitting function. We then construct a histogram weighted by the Akaike information criterion and we take the standard deviation of this distribution as the systematic error of the fit.

\begin{figure}[b]
    \centering
   \includegraphics{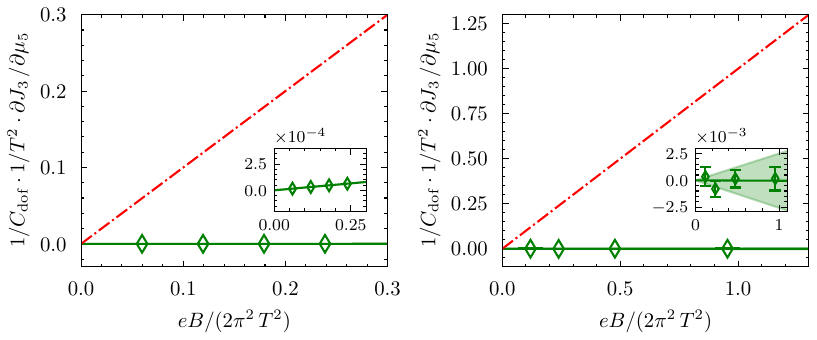}
          \caption{Derivative of the vector current with respect to the chiral chemical potential as a function of the magnetic field for a $24^3 \times 6$ lattice with staggered quarks in the free case with $m/T=1$ (left) and in full QCD with $2+1$ flavors and physical quark masses at $T=305$ MeV (right). The continuous green line represents the optimal fit, and its width is our estimate for its uncertainty. For comparison, the dashed-dotted red line indicates a linear function with unit slope (notice the rescaling of the horizontal axis). For free fermions, the data points were obtained using exact diagonalization (see App.~\ref{sec: free appendix}) and thus have no errors. A nonzero slope, suppressed by four orders of magnitude, is due to residual finite volume effects and lattice artifacts. In the interacting case, the uncertainty of the linear fit is obtained by adding statistical and systematical errors in quadrature.}
            \label{fig:fit}
\end{figure}

In Fig.~\ref{fig:fit} we display two examples of the behavior of the current derivative with $eB$ and the fits, both for free fermions and full QCD with staggered quarks. In these plots, we can see that the expected vanishing result is completely reproduced with staggered fermions. For a complete understanding of CME on the lattice with different discretizations and setups, we now present a detailed analysis of the obtained results for $\CME$ in a wide range of situations.

\section{Results} \label{sec:Results}

\subsection{Free quarks} \label{subsec:free quarks}
A first simple case where we can check our setup is turning off gluonic interactions. We consider a color-singlet theory of free quarks, with mass $m$ and charge $q$. In this particular case, the overall normalization factor reduces to $C_{\rm dof}=(q/e)^2$. 

\begin{figure}[ht]
    \centering
    \includegraphics{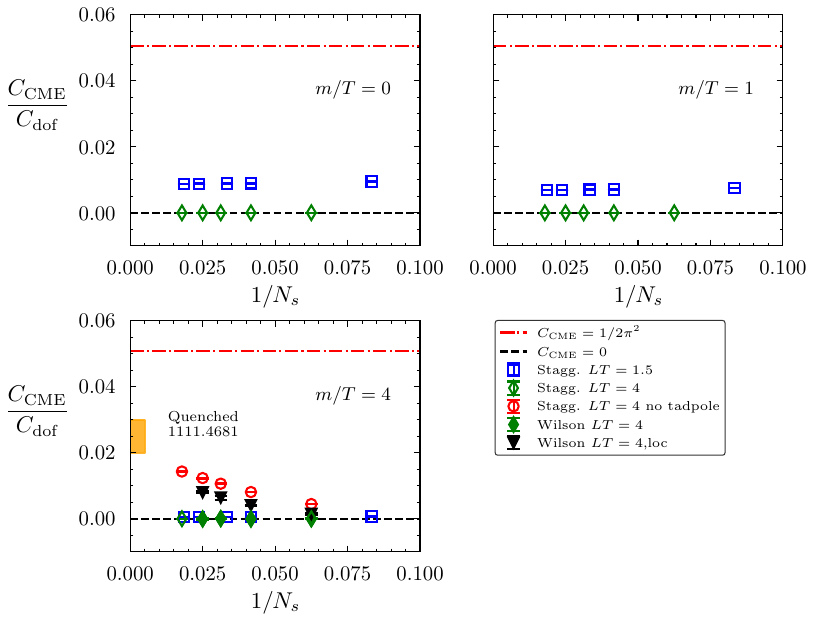}
    \caption{Continuum extrapolation of the CME conductivity for free fermions at different values of $m/T$ using the staggered discretization. 
    The red dot-dashed line corresponds to $\CME=1/(2\pi^2)$, while the black dashed line to $\CME=0$. 
    For light fermions ($m/T=0$ and $m/T=1$, upper panels), the CME coefficient is nonzero on the smallest volume (see the blue points with $LT=1.5$) due to finite size effects and approaches zero in the thermodynamic limit (see the open green symbols for the $LT=4$ results, already shown in the left panel of Fig.~\ref{fig:fit}). For $m/T=4$, Wilson fermions as well as discretizations with a non-conserved vector current are also included. The quenched continuum limit estimation with Wilson fermions from Ref.~\cite{Yamamoto:2011ks} is depicted as an orange band for comparison. }
            \label{fig:free}
\end{figure}

In Fig.~\ref{fig:free}, we present the results for $\CME$ at three different values of $m/T$ with staggered fermions. For the highest $m/T$, we also include a comparison to Wilson fermions. In each case, we show continuum extrapolations ($N_s\to\infty$ with $m/T$ and $LT$ kept fixed) for different aspect ratios $LT$. In small volumes (see the results for $LT=1.5$), we find small values of $\CME$ to remain in the continuum limit. However, these turn out to be finite-size effects, which vanish towards the thermodynamic limit, where Bloch's theorem, ensuring $\CME=0$, becomes valid. In particular, we find $LT=4$ to be sufficiently close to the infinite volume limit for all masses, and to give results for $\CME$ smaller than $10^{-6}$.

In this system of free fermions, we already learn what is the main conclusion of the following results: CME vanishes on a Euclidean lattice if a conserved vector current and an anomalous axial current are considered, both for staggered and Wilson fermions. For $m/T=4$, we also show what happens when this correct setup is abandoned. For staggered quarks, excluding the tadpole terms leads to a deviation of the expected vanishing result in the continuum limit. The same happens for Wilson fermions if a local vector current is used instead of a conserved one. This is exactly the setup used in Ref.~\cite{Yamamoto:2011ks}, whose result for the quenched theory are marked by an orange band in the bottom left plot. With this free fermion calculation we see a first indication that the reason behind the non-zero result in that work could be closely related to the use of a local vector current. This will be confirmed next, by performing QCD simulations in the quenched theory.

\begin{figure}[ht]
    \centering
    \includegraphics{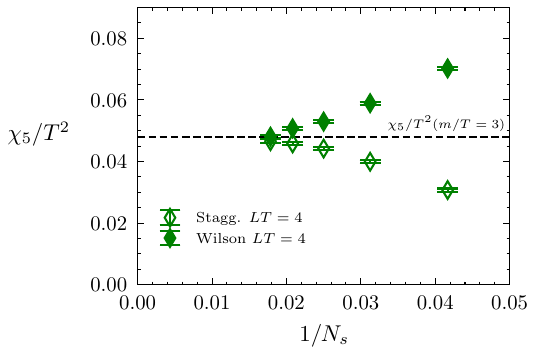}
    \caption{Axial susceptibility for free fermions with $m/T=3$, using both the staggered and the Wilson discretizations. The black dashed line represents the value from the analytical expression~\eqref{eq:chi5analy} at this particular value of $m/T$.} 
    \label{fig:chi5_free}
\end{figure}

Before continuing with interacting theory, there is another important crosscheck that can be done. In Fig.~\ref{fig:chi5_free}, we present results for $\chi_5$ for staggered and Wilson fermions with $m/T=3$, where the required traces have been estimated with Gaussian random vectors in the two cases. For both discretizations, the continuum extrapolation agrees with the analytic calculation (\ref{eq:chi5analy}). This clearly shows that chiral density is non-zero at finite $\mu_5$ in our setup, illustrating that the vanishing result for CME arises from a non-trivial cancellation, and not from the absence of chiral imbalance. 

\subsection{Quenched theory} \label{subsec:quenched}

As the next step towards the full QCD theory, we present our results in the quenched setup. This theory not only allows us to have a first look at how interactions affect the picture, but also enables a direct comparison to existing literature, in particular to the non-vanishing results in Ref.~ \cite{Yamamoto:2011ks}.

\begin{figure}[b]
    \centering
    \includegraphics{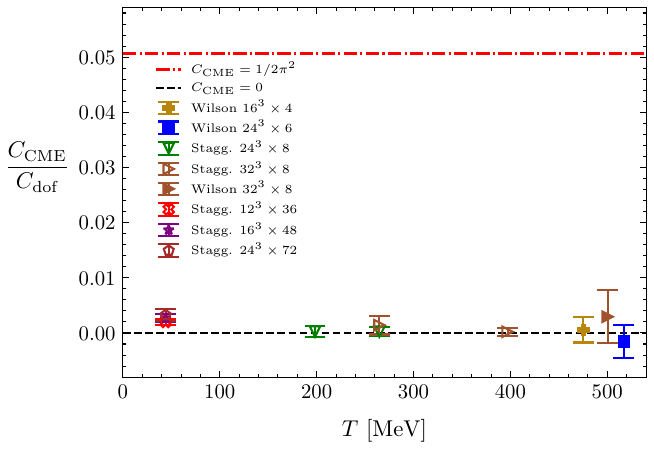}
    \caption{Results for the CME conductivity in the quenched theory, both with Wilson and staggered fermions. The conductivity is consistent with zero in all the cases (within $2\sigma$) even without taking the continuum limit.}
    \label{fig:qnch}
\end{figure}

In Fig.~\ref{fig:qnch}, we present the calculation of $\CME$ in the quenched approximation. We use configurations generated with the plaquette gauge action at $\beta=5.845, 5.9, 6.0, 6.25, 6.26$ and $6.47$, and use both staggered and Wilson fermions in the valence sector. 
We employed these gauge configurations already in Refs.~\cite{Bali:2017ian,Bali:2018sey}. The latter choice is motivated to compare to Ref.~\cite{Yamamoto:2011ks}. The correlator is calculated using a conserved vector current and an anomalous axial current for both discretizations, and the result clearly indicates a vanishing CME.

\begin{figure}[ht]
    \centering
    \includegraphics{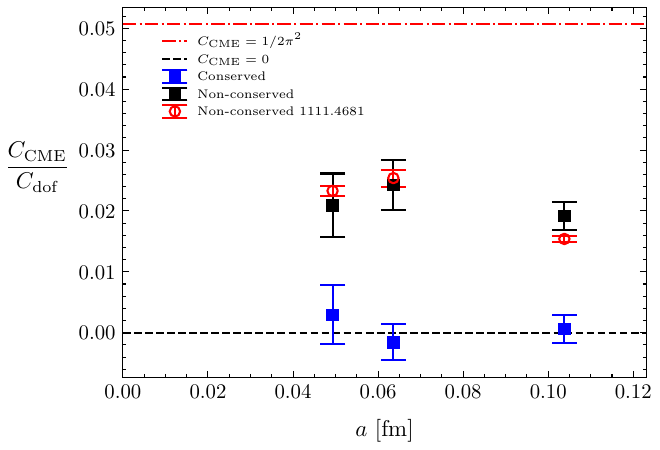}
    \caption{Direct comparison of the study in \cite{Yamamoto:2011ks} (open red circles), with our setup with a conserved vector current (filled blue squares) and a non-conserved one (filled black squares). The results with non-conserved currents agree with each other within errors and deviate from the correct value. The result with a conserved vector current is consistent with a vanishing $\CME$.}
    \label{fig:qnch_yama}
\end{figure}

As mentioned above, we can compare the quenched results from~\cite{Yamamoto:2011ks} by simulating directly with the same parameters, but with a different approach, since the results in this work were obtained by simulating at finite chiral chemical potential and with a local vector current. In Fig.~\ref{fig:qnch_yama}, we present the results for Wilson fermions in the quenched theory, with a conserved vector current and with a local one. The results with the conserved current are consistent with a vanishing $\CME$, while the ones with the local vector current are non-zero and in agreement with the results of Ref.~\cite{Yamamoto:2011ks}. At this point we can conclude that the conservation of the vector current is crucial to reproduce the expected physical picture of a vanishing CME in equilibrium, and the non-zero results for $\CME$ in Ref.~\cite{Yamamoto:2011ks} are an artifact of using a non-conserved vector current.

Finally, we note that in the quenched ensembles gluon configurations in different Polyakov loop sectors contribute. Contrary to the case of the CSE in the quenched approximation~\cite{Brandt:2023wgf}, the Polyakov loop backgrounds do not affect the CME conductivity appreciably. This is because the complex Polyakov loop backgrounds effectively amount to an imaginary baryon chemical potential, and therefore only affect the baryon density (entering the CSE) but not the chiral density.

\subsection{QCD at physical quark masses} \label{subsec: full qcd}
Finally, we present our results for $\CME$ in full QCD using $2+1$ flavors of staggered fermions with physical quark masses. The measurements were performed on an already existing ensemble of configurations for different magnetic fields~\cite{Bali:2011qj,Bali:2012zg}.
 
\begin{figure}[ht]
    \centering
    \includegraphics{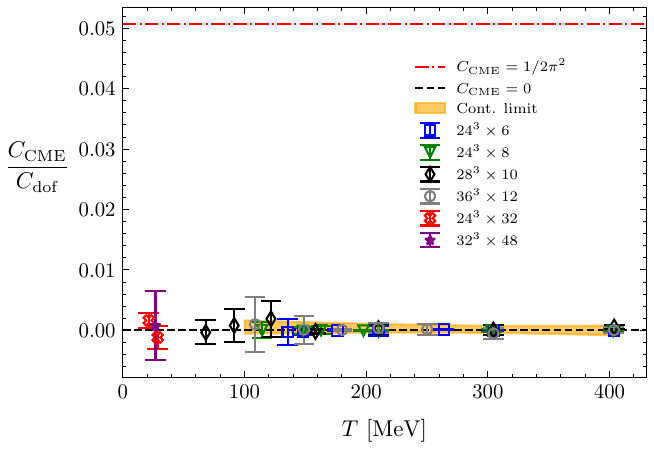}
    \caption{Full dynamic QCD results with $2+1$ flavors of staggered fermions at the physical point in a wide range of temperatures. The continuum limit for points at $T>100$ MeV is shown as an orange band.}
    \label{fig:full_qcd}
\end{figure}

In Fig.~\ref{fig:full_qcd}, we present the dependence of the conductivity on the temperature using several finite-temperature lattice ensembles $24^3 \times 6$, $24^3 \times 8$, $28^3 \times 10$, $36^3 \times 12$ as well as two zero-temperature ensembles $24^3 \times 32$ and $32^3 \times 48$. The CME coefficient vanishes for all temperatures within errors. We also present the continuum limit for temperatures from $100$ MeV to $400$ MeV, considering a spline fit procedure combined with the continuum extrapolation. In particular, we consider a $T$-dependent spline fit of all lattice points with $a$-dependent coefficients of order $a^2$. By minimizing the $\chi^2/\text{dof}$, the optimal fitting surface in the $a-T$ plane is found (see~\cite{Endrodi:2010ai} for further details). The statistical error of this procedure is calculated using the jackknife samples, while we estimate the systematic error by repeating several times the spline fit removing the coarsest lattice and including $\mathcal{O}(a^4)$ coefficients. The maximum difference between the continuum limits obtained with these data sets and the original one is taken as the systematic error, which is added in quadrature to the statistical one. The so-obtained continuum extrapolation is again compatible with zero at all temperatures.

We mention that both the chiral chemical potential and the vector current may be defined with different flavor quantum numbers, i.e.\ baryon number, electric charge or ``light baryon number'', cf.\ Ref.~\cite{Brandt:2023wgf}. We checked that $\CME$ vanishes for any combination of these as well. This is expected, since the general argument implying that CME vanishes in equilibrium, applies in all the different setups. We note that the absence of the CME in certain interacting systems has already been shown in the literature~\cite{Banerjee:2021vvn}. Our result demonstrates that the CME vanishes in equilibrium in full QCD. In addition, it shows that the implications of Bloch's theorem also apply to finite temperature QCD.

\begin{figure}[ht]
    \centering
    \includegraphics{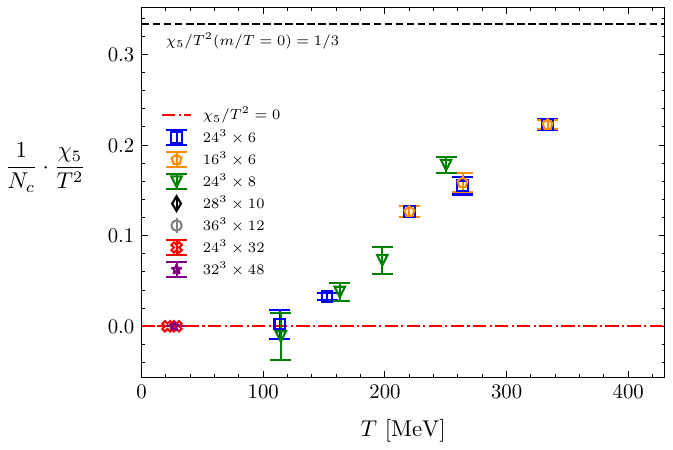}
    \caption{Axial susceptibility for $2+1$ flavors of staggered quarks in full QCD at the physical point. The results are normalized by the number of colors $N_c=3$ in order to directly compare to the free fermions result (\ref{eq:chi5analy}). We observe $\chi_5$ to slowly grow above the crossover temperature, approaching the analytic result for massless non-interacting fermions (indicated by the black dashed line). }
    \label{fig:chi5_full}
\end{figure}

Finally, in Fig.~\ref{fig:chi5_full}, we show the results for the chiral susceptibility $\chi_5$ in full QCD for different temperatures. Since the renormalization\footnote{We note here that $\chi_5$ is also subject to multiplicative renormalization, in particular by the same factor (squared) as the flavor-singlet axial vector current in the CSE \cite{Brandt:2023wgf}. This renormalization constant approaches unity both for Wilson and for staggered fermions in the continuum limit~\cite{Constantinou:2016ieh,Bali:2021qem} and it could be used to further improve the scaling of the observable.} of the observable requires a zero temperature subtraction, approaching the continuum limit is computationally more challenging than for $\CME$. For this reason, we only present results for $N_t=6, 8$ lattices with different volumes. To reduce cutoff effects, we follow a similar improvement program for the observable as in Ref.~\cite{Borsanyi:2010cj}. We multiply our results by the improvement coefficients, which can be calculated for very high temperatures (free quarks with $m/T\approx0$) for the different values of $N_t$. In particular, we find these coefficients to be 1.2505 for $N_t=6$ and 1.1196 for $N_t=8$. The susceptibility starts growing around the crossover temperature $T_c$, and it slowly approaches the analytic result for massless free fermions. The asymptotic limit seems to be reached at temperatures higher than 1 GeV, a similar behavior as observed in the QCD pressure~\cite{Borsanyi:2010cj}, and in contrast with the fast increase after $T_c$ observed in the CSE conductivity~\cite{Brandt:2023dir}.

\section{Summary and outlook} \label{Summary and Outlook}
In this paper, we have presented a complete characterization of the absence of the chiral magnetic effect in thermal equilibrium. We carefully showed how regularization plays a crucial role in the calculation, in particular, for the first-order Taylor coefficient of the conductivity $\CME$. We also reviewed how chiral imbalance is indeed induced by a chiral chemical potential $\mu_5$, even when this parameter does not couple to a conserved quantity in the theory. 

By using first-principles Euclidean lattice calculations, we showed how using the appropriate discretization of the conserved currents leads to a vanishing CME conductivity for non-interacting fermions, using both Wilson and staggered quarks. At the same time chirality is indeed present in the system with both discretizations, indicated by a non-vanishing $\chi_5$. We then moved to the quenched theory, where we again observed the absence of the CME. We also directly compared to the earlier results of Ref.~\cite{Yamamoto:2011ks}, where a non-vanishing CME conductivity has been found, albeit with a non-conserved vector current. The comparison shows that the use of a conserved vector current is vital to obtain consistent results. Finally, we presented the first results in full dynamical QCD at the physical point, using $2+1$ flavors of staggered fermions, for a wide range of temperatures. These results, which have been continuum extrapolated, are completely consistent with $\CME=0$. In addition, we calculated $\chi_5$ in the same setup, exhibiting an increase above the QCD crossover temperature, slowly approaching the non-interacting quarks limit.

Once the equilibrium effect is completely understood, it is natural to extend this study to the out-of-equilibrium conductivity. To access the latter, we would need information about the spectral function, which is accessible via lattice simulations using spectral reconstruction methods. This technique has already been employed to indirectly study the CME in Ref.~\cite{Astrakhantsev:2019zkr} as a contribution to the electric conductivity. Using the temporal correlator between vector and axial vector currents one would have access to the out-of-equilibrium $\CME$ in QCD, which potentially provides information on how the out-of-equilibrium CME is affected by interactions. Important steps in this direction have been taken in Ref.~\cite{Buividovich:2024bmu}, where the role of the regularization and the relation between Euclidean and retarded correlators has been discussed, as well as in Ref.~\cite{Banerjee:2022snd}, where the out-of equilibrium properties of CME for free fermions were analyzed using the Wigner-Weyl formalism.

\acknowledgments
This research was funded by the DFG (Collaborative Research Center CRC-TR 211 ``Strong-interaction matter under
extreme conditions'' - project number 315477589 - TRR 211) and by the Helmholtz Graduate School for Hadron and Ion Research (HGS-HIRe for FAIR). The authors are grateful for inspiring discussions with Pavel Buividovich, Kenji Fukushima, Keh-Fei Liu, Ho-Ung Yee, Dirk Rischke, S\"oren Schlichting, Ilya Selyuzhenkov, Igor Shovkovy and Lorenz von Smekal.  

\appendix

\section{Parameterization dependence of the unregularized $\CME$ in the free case}
\label{app:RegDep}
In this appendix we will analyze the integral appearing in \eqref{eq:cmecancel} in order to show that without an a priori regularization process the CME coefficient is mathematically ill-defined despite not containing UV divergences.
We start by defining two contributions
\begin{align}
\label{eq:fmdef}
    f_m = \frac{1}{4\pi^3}\int_{-\infty}^\infty \dd p_3\, \dd p_4 \frac{m_s^2}{(m_s^2+p_4^2+p_3^2)^2}
\end{align}
and
\begin{align}
\label{eq:fpdef}
    f_p=\frac{1}{4\pi^3}\int_{-\infty}^\infty \dd p_3 \,\dd p_4 \frac{p_4^2-p_3^2}{(m_s^2+p_4^2+p_3^2)^2}\,,
\end{align}
with which \eqref{eq:cmecancel} is just
\begin{align}
    \CME=\sum_{s=0}^{3} c_s(f_m+f_p)\,.
\end{align}

The first term, $f_m$, is a well behaved integral with the value
\begin{equation}
f_m=\frac{1}{4\pi^2}\,.
\end{equation}
In turn, $f_p$ depends on the way one parameterizes the integral over the $p_3-p_4$ plane. First of all, notice that performing the $p_3$ and $p_4$ integrals in a different order gives different results,
\begin{align}
    f_{p,34}=\frac{1}{4\pi^3}\int_{-\infty}^\infty \dd p_3 \left(\int_{-\infty}^\infty \dd p_4 \frac{p_4^2-p_3^2}{(m_s^2+p_4^2+p_3^2)^2}\right)=\frac{1}{4\pi^3}\int_{-\infty}^\infty \dd p_3 \frac{\pi}{2(m_s^2+p_3^2)^{3/2}} = \frac{1}{4\pi^2}\,,
    \label{eq:fp34}
\end{align}
while
\begin{align}
    f_{p,43}=\frac{1}{4\pi^3}\int_{-\infty}^\infty \dd p_4 \left(\int_{-\infty}^\infty \dd p_3 \frac{p_4^2-p_3^2}{(m_s^2+p_4^2+p_3^2)^2}\right)=\frac{1}{4\pi^3}\int_{-\infty}^\infty \dd p_4 \frac{-\pi}{2(m_s^2+p_4^2)^{3/2}} = -\frac{1}{4\pi^2}\,.
    \label{eq:fp43}
\end{align}
More generally, one can rotate the integration variables 
\begin{align}
    \begin{pmatrix}
    p_3\\
    p_4
    \end{pmatrix} = \begin{pmatrix}
    \cos \alpha & \sin\alpha\\
    -\sin\alpha & \cos\alpha
    \end{pmatrix}\begin{pmatrix}
    u\\
    v
    \end{pmatrix}
\end{align}
to find that depending on the (arbitrary) rotation angle $\alpha$, the result changes continuously between the two extremes~\eqref{eq:fp34} and~\eqref{eq:fp43},
\begin{align}
\label{eq:uv}
    f_{p,\alpha}=-\frac{\cos 2\alpha}{4\pi^2}\,.
\end{align}
Taking for example $\alpha=\pi/2$, i.e.\ considering the ordering $f_{p,34}$, results in the value $1/(2\pi^2)$ for the unregularized contribution of the $s=0$ field to $\CME$.

\begin{figure}[ht]
    \centering
    \includegraphics[width=9cm]{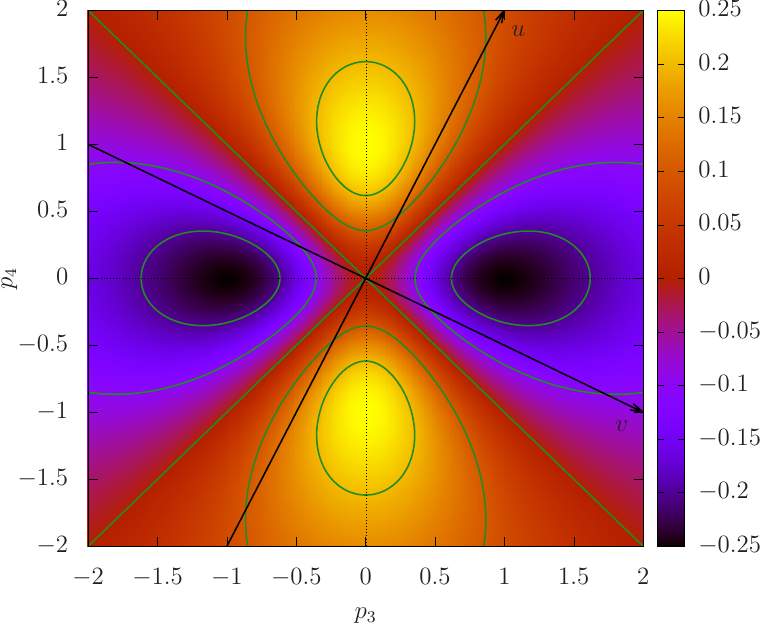}
    \caption{The heat map  of the integrand of $f_p$ as defined in \eqref{eq:fpdef}. A few contours are highlighted in green for better visibility, while an example of the variable change in~\eqref{eq:uv} is shown by the rotated coordinate system (black).
    }
    \label{fig:fp_intgand}
\end{figure}

The heat map of the integrand of $f_p$ for $m_s=1$ is shown in Fig.~\ref{fig:fp_intgand}, with an example change of coordinates, illustrating how the different results arise.
To understand the origin of this ambiguity, it is instructive to use polar coordinates $(p,\varphi)$, writing
\begin{align}
    f_p = \frac{1}{4\pi^3} \int_0^\infty \dd p \int_0^{2\pi} \dd \varphi \,\frac{p^3 \cos 2\varphi}{(m_s^2 + p^2)^2}\,,
\end{align}
which highlights the fact that there are divergent directions (in fact almost all of them are), since the integrand only falls off as $p^{-1}$ for large $p$. However, integrating over $\varphi$ combines them in such a way that cancellations occur to make it possibly finite. This completes our illustration of the ill-definedness of $f_p$ and hence the unregularized value of $\CME$.

As a final remark, evaluating the sum over the PV fields always leads to a well defined integral, which is easiest to see for the form of $f_p$ in polar coordinates, where the integrand falls off faster for $p\gg m_s$ owing to the PV subtraction,
\begin{align}
    \sum_{s=0}^3 c_s f_p &= \frac{1}{4\pi^3} \int_0^\infty \dd p \int_0^{2\pi} \dd \varphi \sum_{s=0}^3 c_s \frac{p^3 \cos 2\varphi}{(m_s^2 + p^2)^2} \\
    &=\frac{1}{4\pi^3} \int_0^\infty \dd p \int_0^{2\pi} \dd \varphi \left(3\frac{m^4+m_1^4-2m_2^4}{p^5}+{\cal O}(p^{-7})\right)\cos 2\varphi\,,
\end{align}
and vanishes due to the $\varphi$-integral. Here we inserted the specific PV masses~\eqref{eq:PVmasses}.

\section{CME in the free case from the two-point function at finite temperature}
\label{sec:cmefinT}

In this appendix, we generalize the calculation of Sec.~\ref{sec:cmefree} to finite temperatures $T$. This is done in a straightforward manner, by replacing the frequency component of the integral by a sum over fermionic Matsubara frequencies,
\begin{align}
	\int \frac{\dd p_0}{2\pi} \to i T\sum_{n=-\infty}^\infty\,,
\end{align}
while also replacing $p_0\to i \omega_n= i 2\pi T(n+1/2)$ in $\widetilde S_s(p)$. Doing so in Eq.~\eqref{eq:cme_p3z1z2},  yields
\begin{align}
\CME q B = -\frac{qB}{2\pi^2}\sum_{s=0}^{3} c_s T \sum_{n=-\infty}^\infty \int_{-\infty}^\infty \dd p_3\int_0^\infty  \dd Z_1 \dd Z_2 \,e^{-(Z_1+Z_2)(\omega_n^2 + m_s^2+p_3^2)}
	\left(m_s^2+\omega_n^2-p_3^2\right)\,.\label{eq:cme_mag_fact}
\end{align}
The linear magnetic field dependence is again obtained, which allows the simplification of the expression. Doing the $p_3$ integral results in
\begin{align}
\CME = -\frac{T}{2\pi^{3/2}}\sum_{s=0}^{3} c_s \int_0^\infty \dd Z_1 \dd Z_2 \sum_{n=-\infty}^\infty\, e^{-(Z_1+Z_2)(\omega_n^2 + m_s^2)}\left[\frac{m_s^2+\omega_n^2}{\sqrt{Z_1+Z_2}}-\frac{1}{2(Z_1+Z_2)^{3/2}}\right]\,.
\end{align}
The $Z_1,Z_2$ integrals can be performed for all terms of the sum, e.g. by changing variables to $Z_1=\frac{\alpha+\beta}{2}$ and $Z_2=\frac{\alpha-\beta}{2}$. The final result is again zero, now for all temperatures.

\section{Axial susceptibility in the free case}
\label{sec:freechi5}
In this appendix we calculate the axial susceptibility~\eqref{eq:axialsusc} in the free case.
First, we discuss $\chi_5$ at $T=0$ in order to illustrate that this observable is subject to additive renormalization. We consider the second derivative of the partition function with respect to $\mu_5$ again in the PV regularization,
\begin{align}
   \chi^b_5=i\sum_{s=0}^3c_s\int \dfrac{\dd^4k}{(2\pi)^4} \dfrac{\Tr[\gamma_0\gamma_5(\slashed{k}+m_s)\gamma_0\gamma_5(\slashed{k}+m_s)]}{(k^2-m^2_s)^2}\,.
   \label{eq:B1}
\end{align}
Carrying out the traces and using the Schwinger parameterization 
\begin{align}
   \chi^b_5=4i\sum_{s=0}^3c_s\int^\infty_0\dd t \, \int \dfrac{\dd^4k}{(2\pi)^4} (2k_0^2-k^2-m^2)\,t\, e^{-t(m_s^2-k^2)}\,.
\end{align}
The momentum integrals can be solved, yielding
\begin{align}
   \chi^b_5=\dfrac{1}{4\pi^2}\sum_{s=0}^3c_s\int^\infty_0\dd t\, \qty[-\frac{1}{t^2}+\frac{m_s^2}{t}] e^{-t(m_s^2-k^2)}\,.
\end{align}
This integral is divergent for the physical species $s=0$, but it is regularized by the PV copies. The final result after integration reads
\begin{align}
   \chi^b_5=-\dfrac{1}{2\pi^2}\sum_{s=0}^3c_s\,m_s^2\log\qty(\frac{m_s^2}{m^2})=-\dfrac{1}{2\pi^2}\qty[\log(4)\Lambda^2-m^2\qty[1+\log\qty(\frac{\Lambda^2}{2m^2})]+\mathcal{O}(\Lambda^{-2})]\,,
\end{align}
where we inserted the specific form~\eqref{eq:PVmasses} of the PV masses, noting that the physical mass is $m=m_0$.
Here we explicitly see how the introduction of $\mu_5$ leads to a quadratic divergence (plus a logarithmic one) in $\log \mathcal{Z}$. This motivates the $T=0$ subtraction of the bare observable in~\eqref{eq:chi5sub}.

To calculate the finite temperature result, is again enough to use the Matsubara formalism. From Eq.~\eqref{eq:B1} and rotating to imaginary time, we get
\begin{align}
   \chi^{T}_5=-4\, T \sum_{n=-\infty}^\infty\int \dfrac{\dd^3k}{(2\pi)^3} \dfrac{(i\omega_n)^2+\vec{k\,}^2-m^2}{[(i\omega_n)^2-\Vec{k\,}^2-m^2]^2}\,.
\end{align}
It is straightforward to evaluate the Matsubara sum, yielding, for the $T$-dependent part,
\begin{equation}
    \chi_5=\dfrac{4}{\pi^2} \int \dd k \, \dfrac{k^2}{E_k}n_F(E_k)\,,
\end{equation}
which is Eq.~\eqref{eq:chi5analy}.

\section{CME with free staggered fermions}
\label{sec: free appendix}

In this appendix, we illustrate how to calculate $\CME$ for free staggered quarks using explicitly the eigenvalues and eigenvectors of the Dirac operator. The discussion closely follows the analogous method that we developed for the CSE~\cite{Brandt:2023wgf}. To describe free quarks, we turn off the gluon fields and consider one color component only. We discuss a single quark flavor with mass $m$, the chemical potential $\mu_5$ and the vector current $J_{3}$. Below we work on an $N_s^3\times N_t$ lattice and write everything in lattice units, setting $a=1$. In the non-interacting theory, the disconnected term of Eq.~\eqref{eq:derstag} vanishes, so we only need to calculate two terms
\begin{equation}
   \eval{\pdv{J_{3}}{\mu_5}}_{\mu_5=0}=\dfrac{1}{N_s^3N_t}\Bigg[-\dfrac{1}{4}\text{Tr}\qty(\Gamma_3 M^{-1}\Gamma_{45} M^{-1})
    +\dfrac{1}{4}\text{Tr}\qty(\dfrac{\partial\Gamma_{3}}{\partial \mu_5}M^{-1})\Bigg]\,,\label{eq:freederstag}
\end{equation}
and the expectation value indicating the fermion path integral was omitted for brevity.

The massless staggered Dirac operator has purely imaginary eigenvalues, due to its antihermiticity. Moreover, due to staggered chiral symmetry, $\{\slashed{D},\eta_5\}=0$ (with $\eta_5=(-1)^{n_1+n_2+n_3+n_4}$), the eigenvalues come in complex conjugate pairs,
\begin{equation}
    \slashed{D}\Psi_i=\pm i\lambda_i \Psi_i\,.
    \label{eq:B2}
\end{equation}
When the mass is included, we will need the analogous eigensystem for $M M^\dagger = (\slashed{D}+m) (\slashed{D}+m)^\dagger = \slashed{D}\slashed{D}^\dagger+m^2$, so
\begin{equation}
    M M^\dagger\Psi_i=(\lambda_i^2+m^2) \Psi_i \,,
    \label{eq:B3}
\end{equation}
where each eigenvalue is doubly degenerate due to~\eqref{eq:B2}.
Using this eigensystem as basis, the traces in Eq.~\eqref{eq:freederstag} can be written as 
\begin{equation}
\begin{split}
   \eval{\pdv{J_{3}}{\mu_5}}_{\mu_5=0}=
    \dfrac{1}{N_s^3N_t}\Bigg[&-\dfrac{1}{4}\sum_{i,j}\dfrac{1}{(\lambda^2_i+m^2)(\lambda^2_j+m^2)}\Psi^\dagger_i \Gamma_3 M^\dagger \Psi_j\Psi^\dagger_j\Gamma_{45}M^\dagger \Psi_i\\
    &+\dfrac{1}{4}\sum_i\dfrac{1}{\lambda^2_i+m^2}\Psi^\dagger_i\dfrac{\partial\Gamma_{3}}{\partial \mu_5}M^\dagger\Psi_i\Bigg]\,,
    \end{split}
    \label{eq:spectralsum}
\end{equation}
where we inserted a complete set of eigenstates $\sum_j\Psi_j \Psi_j^\dagger=\mathds{1}$ in the first term and used $M^{-1}=M^\dagger(MM^\dagger)^{-1}$.

Next, the separability of the problem allows us to reduce Eq.~\eqref{eq:B3} to one two-dimensional and two one-dimensional eigenproblems. This will enable us to determine the complete spectrum on much larger lattices than with a direct, four-dimensional approach. To this end, we write the free Dirac operator (at $\mu_5=0$) as $\slashed{D}=\slashed{D}_{12}+\slashed{D}_3+\slashed{D}_4$ with $\slashed{D}_{12}=\slashed{D}_1+\slashed{D}_2$ and
\begin{equation}
\slashed{D}_{\nu}(n,m)=\dfrac{\eta_\nu(n)}{2} \left[u_\nu(n)\delta_{n+\hat{\nu},m}-u^*_\nu(n-\hat{\nu})\delta_{n-\hat{\nu},m}\right]\equiv \eta_\nu(n)D_\nu(n,m)\,.
\end{equation}
Similarly, the staggered Dirac matrices~\eqref{eq:gammas} (at $\mu_5=0$) simplify to
\begin{align}
    \Gamma_\nu(n,m)=\dfrac{\eta_\nu(n)}{2} \left[u_\nu(n)\delta_{n+\hat{\nu},m}+u^*_\nu(n-\hat{\nu})\delta_{n-\hat{\nu},m}\right]\equiv \eta_\nu(n)S_\nu(n,m)\,.
\end{align}
Notice that the hop operators satisfy 
\begin{equation}
    [S_1,S_3]= [S_1,S_4]=[S_2,S_3]=[S_2,S_4]=[S_3,S_4]=0\,,
    \label{eq:Scomm}
\end{equation}
because the $\mathrm{U}(1)$ links only enter in $S_1$ and $S_2$.

The squared operator $MM^\dagger=MM^\dagger_{12}+MM^\dagger_3+MM^\dagger_4$ separates into three terms that act in the respective subspaces and only depend on the respective coordinates -- thus they commute with each other.
Therefore, the eigenvectors in~\eqref{eq:B3} factorize,
\begin{equation}
    \Psi_{\{i_{12},i_3,i_4\}}(n_1,n_2,n_3,n_4)=\rho_{i_{12}}(n_1,n_2) \,\phi_{i_3}(n_3)\, \xi_{i_4}(n_4)\,,
    \label{eq:factmode}
\end{equation}
with $0\le i_{12}<N_s^2$, $0\le i_3<N_s$ and $0\le i_4<N_t$.\footnote{Note that the same separation does not hold for the eigensystem~\eqref{eq:B2}, since for example $[\slashed{D}_{12},\slashed{D}_3]\neq0$ due to the staggered phases. This is because $[MM^\dagger,\eta_5]=0$ but $[\slashed{D},\eta_5]\neq0$.} Below we use a shorthand notation and simply write $\Psi_i=\rho_i \phi_i \xi_i$. 

We proceed by expanding the operators appearing in the matrix elements in Eq.~\eqref{eq:spectralsum} into separate contributions that depend only on $n_{1,2}$, $n_3$ or $n_4$,
\begin{align}
    \Gamma_3&=
    (-1)^{n_1+n_2}S_3\,,\\
    M^\dagger&=-\slashed{D}_{12}-(-1)^{n_1+n_2} D_3-(-1)^{n_1+n_2+n_3}D_4+m\,,
\end{align}
and it can be shown with some algebra and Eq.~\eqref{eq:Scomm} that
\begin{equation}
    \Gamma_{45}= -S_{12}S_3 (-1)^{n_2}\,,
\end{equation}
with $S_{12}=\{S_1,S_2\}/2$.
Finally, for the tadpole term we need the derivative of $\Gamma_{3}$ with respect to the chiral chemical potential. Using \eqref{eq:gammas} and \eqref{eq:hmu5},
\begin{align}
        \dfrac{\partial\Gamma_{3}}{\partial \mu_5}=-S_{12}D_3 (-1)^{n_2}\,.
\end{align}

The necessary products, appearing in~\eqref{eq:spectralsum} are therefore
\begin{equation}
\begin{split}
        \Gamma_3 M^\dagger=-&[(-1)^{n_1+n_2}\slashed{D}_{12}]_{12}\cdot[S_3]_3\cdot[\mathds{1}]_4\\
        -&[\mathds{1}]_{12}\cdot[S_3D_3]_3\cdot[\mathds{1}]_4\\
        -&[\mathds{1}]_{12}\cdot[S_3(-1)^{n_3}]_3\cdot[D_4]_4\\
        +&m[(-1)^{n_1+n_2}]_{12}\cdot[S_3]_3\cdot[\mathds{1}]_4\,,
\end{split}
\end{equation}
where $[.]_{12}$, $[.]_3$ and $[.]_4$ indicate operators that only act in the respective spaces and only depend on the respective coordinates. Similarly, we obtain
\begin{equation}
\begin{split}
        \Gamma_{45} M^\dagger=+&[S_{12}(-1)^{n_2}\slashed{D}_{12}]_{12}\cdot[S_3]_3\cdot[\mathds{1}]_4\\
        +&[S_{12}(-1)^{n_1}]_{12}\cdot[S_3D_3]_3\cdot[\mathds{1}]_4\\
        +&[S_{12}(-1)^{n_1}]_{12}\cdot[S_3(-1)^{n_3}]_3\cdot[D_4]_4\\
        -&m[S_{12}(-1)^{n_2}]_{12}\cdot[S_3]_3\cdot[\mathds{1}]_4\,,
\end{split}
\end{equation}
and 
\begin{equation}
\begin{split}
       \dfrac{\partial \Gamma_{3}}{\partial \mu_5} M^\dagger=&+[S_{12}(-1)^{n_2}\slashed{D}_{12}]_{12}\cdot [D_3 ]_3\cdot[\mathds{1}]_4\\
        &+[S_{12}(-1)^{n_1}]_{12}\cdot[D_3 D_3]_3\cdot[\mathds{1}]_4\\
        &+[S_{12}(-1)^{n_1}]_{12}\cdot[D_3 (-1)^{n_3}]_3\cdot[D_4]_4\\
        &-m[S_{12}(-1)^{n_2}]_{12}\cdot[D_3]_3\cdot[\mathds{1}]_4\,.
\end{split}
\label{eq:B18}
\end{equation}
Combining everything, Eq.~\eqref{eq:spectralsum} becomes
\begin{equation}
\begin{split}
          \dfrac{\partial J_{3}}{\partial \mu_5}\Big|_{\mu_5=0} =+&\dfrac{1}{N_s^3N_t}\dfrac{1}{4}\sum_{i,j}\dfrac{1}{(\lambda^2_i+m^2)(\lambda^2_j+m^2)}\\ &\Big\{[A_{ij}C_{ji}]_{12}\cdot[G_{ij}G_{ji}]_3\cdot[\delta_{ij}]_4+[A_{ij}E_{ji}]_{12}\cdot[G_{ij}H_{ji}]_3\cdot[\delta_{ij}]_4\\
          &+[A_{ij}E_{ji}]_{12}\cdot[G_{ij}I_{ji}]_3\cdot[\delta_{ij}J_{ji}]_4-m[A_{ij}F_{ji}]_{12}\cdot[G_{ij}G_{ji}]_4\cdot[\delta_{ij}]_4\\
          &+[\delta_{ij}C_{ji}]_{12}\cdot[H_{ij}G_{ji}]_3\cdot[\delta_{ij}]_4+[\delta_{ij}E_{ji}]_{12}\cdot[H_{ij}H_{ji}]_4\cdot[\delta_{ij}]_4\\
          &+[\delta_{ij}E_{ji}]_{12}\cdot[H_{ij}I_{ji}]_3\cdot[\delta_{ij}J_{ji}]_4-m[\delta_{ij}F_{ji}]_{12}\cdot[H_{ij}G_{ji}]_4\cdot[\delta_{ij}]_4\\
          &+[\delta_{ij}C_{ji}]_{12}\cdot[I_{ij}G_{ji}]_3\cdot[J_{ij}\delta_{ji}]_4+[\delta_{ij}E_{ji}]_{12}\cdot[I_{ij}H_{ji}]_3\cdot[J_{ij}\delta_{ji}]_4\\
          &+[\delta_{ij}E_{ji}]_{12}\cdot[I_{ij}I_{ji}]_3\cdot[J_{ij}J_{ji}]_4-m[\delta_{ij}F_{ji}]_{12}\cdot[I_{ij}G_{ji}]_3\cdot[J_{ij}\delta_{ij}]_4\\        
          &-m[B_{ij}C_{ji}]_{12}\cdot[G_{ij}G_{ji}]_3\cdot[\delta_{ij}]_4-m[B_{ij}E_{ji}]_{12}\cdot[G_{ij}H_{ji}]_3\cdot[\delta_{ij}]_4\\
          &-m[B_{ij}E_{ji}]_{12}\cdot[G_{ij}I_{ji}]_3\cdot[\delta_{ij}J_{ji}]_4+m^2[B_{ij}F_{ji}]_{12}\cdot[G_{ij}G_{ji}]_3\cdot[\delta_{ij}]_4\Big\}\\
          +&\dfrac{1}{N_s^3N_t}\dfrac{1}{4} \sum_i \dfrac{1}{\lambda^2_i+m^2}\\
          &\Big\{[C_{ii}]_{12}\cdot[K_{ii}]_3\cdot[\delta_{ii}]_4
          +[E_{ii}]_{12}\cdot[N_{ii}]_3\cdot[J_{ii}]_4\\
        &+[E_{ii}]_{12}\cdot[L_{ii}]_3\cdot[\delta_{ii}]_4
        -m[F_{ii}]_{12}\cdot[K_{ii}]_3\cdot[\delta_{ii}]_4 \Big\}\,,
     \end{split}
     \label{eq:freefinal}
\end{equation}
with
\begin{equation}
\begin{split}
    &A_{ij} \equiv \rho^\dagger_i(-1)^{n_1+n_2}\slashed{D}_{12} \rho_j\,, \\ 
    &C_{ij}\equiv \rho^\dagger_i S_{12}(-1)^{n_2}\slashed{D}_{12} \rho_j \,, \\
    &F_{ij}\equiv \rho^\dagger_i S_{12}(-1)^{n_2} \rho_j \,, \\
    &G_{ij}\equiv \phi^\dagger_i S_3\phi_j \,, \\
    &I_{ij}\equiv \phi^\dagger_i S_3(-1)^{n_3}\phi_j \,, \\
    &L_{ij}\equiv \phi^\dagger_i D_3 D_3\phi_j \,, 
\end{split}     
\qquad
\begin{split}
        &B_{ij} \equiv \rho^\dagger_i(-1)^{n_1+n_2} \rho_j \,,\\
        &E_{ij}\equiv \rho^\dagger_i S_{12}(-1)^{n_1} \rho_j \,, \\
        &J_{ij}\equiv \xi^\dagger_i D_4\xi_j \,,\\
        &H_{ij}\equiv \phi^\dagger_i S_3D_3\phi_j  \,, \\
        &K_{ij}\equiv \phi^\dagger_i D_3\phi_j \,, \\
        &N_{ij}\equiv \phi^\dagger_i D_3(-1)^{n_3}\phi_j \,.
\end{split}     
\label{eq:B20}
\end{equation}

To solve the eigenvalue problems, a two-dimensional one in the $12$ plane in the presence of the magnetic field and two one-dimensional ones in the $3$ and $4$ directions, we use the LAPACK library. The eigensystem of the former problem gives Hofstadter's butterfly~\cite{Hofstadter:1976zz}, the spectrum of a well-known solid-state physics model. Its importance for lattice QCD has been pointed out in Ref.~\cite{Endrodi:2014vza}, and it has been generalized both for gluonic interactions~\cite{Bruckmann:2017pft,Bignell:2020dze} as well as for inhomogeneous magnetic fields~\cite{Brandt:2023dir}. The value for the current derivative can be reconstructed using Eq.~\eqref{eq:freefinal}, and $\CME$ can be extracted in the same way as explained in the main text by calculating the observable at different values of the magnetic field.

\bibliographystyle{JHEP}
\bibliography{biblio.bib}

\end{document}